\pdfoutput=1

\documentclass[10pt]{article}
\usepackage{graphicx}
\usepackage{appendix}
\usepackage{multicol}
\usepackage{graphicx}
\usepackage{subfigure}
\usepackage{url}
\usepackage{caption}
\usepackage{amsmath}
\usepackage{amsfonts}
\usepackage{amssymb}

\def\newblock{\hskip .11em plus .33em minus .07em}

\begin{document}
\title{Thermodynamics and time-directional invariance.}
\author{A. Y. Klimenko\thanks{Email for communications: klimenko@mech.uq.edu.au  } \\ 
{\it The University of Queensland, SoMME, QLD 4072, Australia} \\ 
\and U.Maas  \\ 
{ \it Karlsruher Institut fur Technologie, ITT, 76131, Germany} }

\maketitle

\begin{abstract}
Time directions are not invariant in conventional thermodynamics. We
investigate implications of postulating time-direction invariance  in
thermodynamics and requiring that thermodynamic descriptions are not changed
under time reversal accompanied by replacement of matter by antimatter (i.e.
CPT-invariant thermodynamics).  The matter and antimatter are defined as
thermodynamic concepts without detailing their physical structure.

Our analysis stays within the limits of conceptual thermodynamics and leads to
effective negative temperatures, to thermodynamic restrictions on time travel
and to inherent antagonism of matter and antimatter. This antagonism is purely
thermodynamic; it explains the difficulty in achieving thermodynamic
equilibrium between matter and antimatter and does not postulate their mutual
annihilation on contact. We believe that the conclusions of this work can be
of interest not only for people researching or teaching thermodynamics but
also for a wider scientific audience. 
\end{abstract}

\section{Introduction}

\subsection{Boltzmann's time hypothesis}

In the 1890s, the kinetic theory of Ludwig Boltzmann, which represents an
important link between thermodynamics and classical mechanics, attracted both
interest and criticism. The criticism was to some extent motivated by doubts
about the atomic (molecular) structure of matter, which were quite persistent
at that time, but also involved a series of very interesting questions about
consistency of the reversibility of classical mechanics with the irreversible
nature of thermodynamics. Some of these questions (i.g. exact physical
mechanism determining the direction of time) are not fully answered even
today. In response to his critics, Boltzmann put forward a number of
hypotheses of remarkable originality and depth
\cite{Boltzmann-nature,Boltzmann-book}. One of these hypotheses relates the
perceived direction of time to the second law of thermodynamics and the
temporal boundary conditions imposed on the universe. The consequence of this
hypothesis is that, given different temporal boundary conditions, time may run
in opposite directions in different parts of the universe. In other words,
entropy can have an increasing trend in some sections of the universe and a
decreasing in the other sections. This notion was introduced in the context of
giant fluctuations occurring in the eternal universe, but the giant
fluctuation hypothesis (which Boltzmann actually attributed to his assistant,
Dr Schuets \cite{Boltzmann-nature}) is the less interesting part of
Boltzmann's analysis, which was designed to have his ideas understood by the
scientific community of the 19th century. At that time, the prevalent
understanding was that the Universe is eternal and any notion of initial
conditions imposed at the beginning of the universe were not likely to be
accepted as scientific.

The essence of Boltzmann's time hypothesis is that the physical nature of time
is direction-symmetric and so is the principal foundation of thermodynamics.
The culprits of the observed irreversibility are the temporal boundary
conditions imposed on matter in the universe or in the observed part of it
(low entropy in the past and high entropy in the future, referring to the
direction of time as we perceive it). The uneven boundary conditions result in
the common trend of matter to have its entropy increase with time.

As the situation stands at present, we still do not know the exact physical
mechanism that links the temporal boundary conditions (i.e. initial and final
conditions) to the time-directional properties of matter. The temporal
boundary conditions alone are not likely to be sufficient for explanation of
the observed time asymmetry. The modern view, which perhaps is best
articulated by Penrose \cite{PenroseBook}, is that the physical laws should
have a very small temporal directional bias, which is too small to be noticed
in classical mechanics but the bias is dramatically amplified by ergodic
mixing of dynamic trajectories (as studied in statistical mechanics). A tiny
deviation from temporal symmetry, presumably related to decoherence of quantum
states during their interaction with matter, causes an avalanche of the phase
space increases and the flow of the thermodynamic time in one direction. The
fundamental physical laws are close to but not fully time symmetric, although
the symmetry is deemed to be exact if the opposite time directions are
associated with matter and antimatter. At this point we refer to Feynman's
theory of antiparticles \cite{Feynman1949}, which treats antiparticles as
particles moving back in time, and to CPT (charge/parity/time) symmetry in
conjunction with the known minor CP violation \cite{PenroseBook}. \ The
present work endeavours to construct a CPT-invariant thermodynamic theory,
which in absence of antimatter is consistent with conventional thermodynamics.
Note that all thermodynamic theories operating with scalar quantities are
parity-invariant, while CT invariance is of most interest here.

\subsection{Thermodynamic antimatter}

While the exact\ mechanism of time and many other associated questions have
yet to be answered\ by physical sciences, we are interested here in
thermodynamical aspects of the Boltzmann time hypothesis. Following modern
undestanding of this hypothesis combined with Feynman theory of antimatter, we
assume that matter from our world, could be brought into thermodynamic contact
with the so-called \textit{thermodynamic antimatter} populating another
section of the universe (one might call it the antiworld) where the time
direction (i.e. the temporal direction of entropy increase) is opposite to
ours. We use the term ''thermodynamic antimatter'' to follow the modern
interpretation of the ideas of Ludwig Boltzmann and define a purely
thermodynamic concept, which in principle may or may not be related to
anti-particles and physical antimatter. Both matter and antimatter are
presumed to be compliant with the first law of thermodynamics and have
equivalent properties in reversible processes. However, matter and antimatter
have opposite directions of irreversibility: isolated antimatter can decrease
but cannot increase its entropy in our time.

Thermodynamic antimatter is a thermodynamic object that is postulated to
comply with the following principles:

\begin{itemize}
\item \textbf{Reversible equivalence.} There is no distinction between matter
and antimatter with respect to the first law of thermodynamics.

\item \textbf{Inverted irreversibility.} Thermodynamically\ isolated
antimatter can increase its entropy only backward in time (unlike any isolated
matter, whose entropy increases forward in time).

\item \textbf{Observational symmetry.} Antimatter and its interactions with
matter are seen (i.e. observed, experimented with or measured) by
antiobservers in exactly the same way as matter and its interactions with
antimatter are seen by observers.
\end{itemize}

Conservation of energy by the reversible equivalence principle also implies
preservation of mass. Hypothetical observers made of antimatter, are called
\textit{antiobservers,} while the term \textit{observers} refers only to us
--- observers made of matter. Properties of matter measured by us (the
observers) and properties of antimatter measured by antiobservers are referred
to as \textit{intrinsic. }The properties of matter and antimatter measured by
observers are referred to as \textit{apparent, }while the properties of matter
and antimatter measured by antiobservers are referred to as
\textit{antiapparent. }

Despite the relative simplicity of our assumption and analysis, our
conclusions involve an inherent thermodynamic hostility of matter and
antimatter and seem to deviate from existing views. The suggested approach is
different from the conventional assignment of the same irreversible
thermodynamic properties to both matter and antimatter implying violations of
the CPT symmetry \cite{PenroseBook}. The competing physical intuitions of CPT
invariance and of conventional thermodynamics has been recently discussed by
Downes et al \cite{CPT2012}. Our approach is also different from thermodynamic
analysis of physical antimatter by Dunning-Davies \cite{Dunning-Davies}, which
is based on Santilli isodualities: our reversible equivalence principle does
not need anti-photons and antigravity that are associated with Santilli
isodualities and remain seen by most physicists as unlikely.\ Our approach
based on thermodynamic antimatter seems to suggest some alternative
astrophysical interpretations. For example, in his fundamental work
\cite{Hawking1976}, Hawking compared thermodynamics of black and white holes
and concluded that black and white holes of the same mass should have the same
temperatures. Our inference is that, assuming that the thermodynamic
antimatter is real and thus should form white holes, the effective temperature
of white holes should be much higher.

While mentioning similarities and differences with a number of existing
physical theories, we should state that our approach is based on
thermodynamics --- it operates with thermodynamic concepts and objects, which
are defined by postulating their thermodynamic properties.

\section{Thermodynamics of interactions of matter and antimatter}

Thermodynamics is based on determining the direction of processes where states
(i.e. macrostates) can be realised by the largest possible number of
microstates (given the constraints imposed on the system) and thus are
overwhelmingly more likely than states encompassing fewer microstates. The
logic of thermodynamics considers what is likely and neglects what is
unlikely. The most likely state is called equilibrium. In conventional
thermodynamics, unlikely states may be set as initial states while the system
tends to move towards its equilibrium as time passes. This is reflected by the
well-known Boltzmann--Planck entropy equation
\begin{equation}
S_{i}=k_{B}\ln(W_{i}) \label{BP}%
\end{equation}
linking the entropy $S_{i}$ in the state $i$ to the number of microstates
$W_{i}$ in this state. The number of microstates $W_{i}$ is further referred
to as the statistical weight of the thermodynamic state $i$. The constant
$k_{B}$ is the Boltzmann constant that rescales very large changes in $W_{i}$
to more manageable quantities of conventional thermodynamics.

The most difficult part in considering time-symmetric behaviour is the
necessity to suspend the causality principle, which is deeply engraved in our
way of thinking. This necessity was clearly stipulated by Price
\cite{PriceBook} who has thoroughly discussed philosophical aspects of the
direction of time. Indeed, conventional thinking, which is based on causality
and allows the influence of the past on the future but not vice versa, is
inherently time-asymmetric and can not be used for considerations that are
neutral with respect to the direction of time. In conventional thermodynamics,
the likeness of the future states is evaluated for a selected time moment in
the future, given fixed conditions specified in the present. This has to be
replaced by direction-neutral evaluation of the likeness of different time
trajectories, consistent with the boundary conditions, which are fixed by
external means, and other physical laws and constraints. Specifically, the
conditions for matter are fixed in the past while conditions for antimatter
are fixed in the future (from our perspective and in the past from the
antiobserver perspective). According to the declared observational symmetry
principle, our analysis should remain symmetric with respect to time: we (i.e.
the observers) see antimatter in exactly the same way as the antiobservers see
our matter.

\subsection{Apparent temperatures}

The temporal boundary conditions for the example shown in Figure \ref{fig1}
are: $U_{m},S_{m}$ are specified for matter at $t=-t_{0}$ and $\bar{t}=t_{0}$
and $\bar{U}_{m},\bar{S}_{m}$ are specified for antimatter at $\bar{t}=-t_{0}$
and $t=t_{0}$. The overbar symbol indicates that the value is antiapparent,
i.e. evaluated from the perspective of an antiobserver, whose time $\bar
{t}=-t$ goes in the opposite direction as compared to our time $t$. A limited
thermodynamic contact of matter and antimatter, allowing for transition of a
small quantity of thermal energy $\delta Q,$ occurs at $t=0$ (and $\bar{t}%
=0$). According to observer the thermal energy $\delta Q$ is transferred from
antimatter to matter as shown by the black solid arrow. According to the
antiobserver, who has the opposite direction of time the same thermal energy
$\delta Q$ is transferred from matter to antimatter as shown by the red dashed
arrow. Heat $\delta Q$ is assumed to be positive when transferred in the
direction shown in Figure \ref{fig1}: from antimatter to matter according to
the observer and from matter to antimatter according to the antiobserver. The
total energy
\begin{equation}
U_{tot}=U_{m}+\left(  \bar{U}_{a}+\delta Q\right)  =\left(  U_{m}+\delta
Q\right)  +\bar{U}_{a}\label{Utot}%
\end{equation}
(evaluated at any constant time $t=-\bar{t}$) is preserved in this example, as
it should since the formulation of the first law of thermodynamics does not
depend on the differences between matter and antimatter due to the postulated
reversible equivalence. The entropy change of matter as observed by us and the
entropy change of antimatter as observed by the antiobserver (these are the
entropies linked to $W$) can be easily evaluated and these changes of
intrinsic entropy are shown in Figure \ref{fig1} for the states m$^{\prime}$
and a$^{\prime}$.

We now evaluate the overall statistical weight $W_{tot}$ that corresponds to
different trajectories that are allowed by the first law of thermodynamics.
The overall state is related to the four sub-states: m, a, m$^{\prime}$ and
a$^{\prime}.$ The overall statistical weight $W_{tot}$ is linked to the
product of the statistical weights of the sub-states $W_{m}W_{m^{\prime}}%
W_{a}W_{a^{\prime}}$ and, according to equation (\ref{BP}), becomes
\begin{equation}
W_{tot}(\delta Q)\sim W_{m}W_{m^{\prime}}W_{a}W_{a^{\prime}}=\exp\left(
\frac{S_{m}+S_{m^{\prime}}+\bar{S}_{a}+\bar{S}_{a^{\prime}}}{k_{B}}\right)
\label{W1}%
\end{equation}
The value $W_{tot}$ depends on $\delta Q$. Note that $S_{m}$ and $\bar{S}_{a}$
are fixed by the boundary conditions and only $S_{m^{\prime}}$ and $\bar
{S}_{a^{\prime}}$ depend on $\delta Q$. In this example, we should place the
time moments $t=-t_{0}$ and $t=+t_{0}$ as far apart as needed to ensure
establishment of equilibriums within matter and antimatter before and after
the interaction, which is used in (\ref{W1}) in form of stochastic
independence of microstates that correspond to the macrostates m and
m$^{\prime}$, a$^{\prime}$ and a. If this was not the case, then not all
microstates of state m$^{\prime}$ may be accessible due to stochastic
correlations with state m (in simple terms the system may not have enough time
to establish its equilibrium state). Assessing the statistical weight of a
continuous trajectory is, generally, a difficult task due to possible
correlations between microstates. Indeed, in conventional thermodynamics, a
system which is initially placed into a non-equilibrium state, is unlikely to
be transferred in the next moment into any microscate that belongs to the
equilibrium state of the system. However, as sufficient time passes by, all
microstates becomes achievable and the system can be found in its equilibrium
state with overwhelming probability. Here, we place the time moments
$t=-t_{0}$ and $t=+t_{0}$ sufficiently far apart so that the system
microstates at these moments do not correlate. Two of the states, m and a, are
fixed by the boundary conditions.

Equation (\ref{W1}) can be simplified through normalising $W_{tot}$ by the
value of $W_{tot}$ at $\delta Q=0$%
\begin{equation}
\frac{W_{tot}(\delta Q)}{W_{tot}(0)}=\exp\left(  \frac{\delta Q}{k_{B}}\left(
\frac{1}{T_{m}}+\frac{1}{\bar{T}_{a}}\right)  \right)  \label{W2}%
\end{equation}
where conventional definitions of the temperature
\begin{equation}
\frac{1}{T_{m}}=\frac{\partial S_{m}}{\partial U_{m}},\;\;\;\frac{1}{\bar
{T}_{a}}=\frac{\partial\bar{S}_{a}}{\partial\bar{U}_{a}}\label{TT}%
\end{equation}
are used. (The quantity $\delta Q$ is assumed to be too small to affect the
temperatures of matter and antimatter, which remain $T_{m}$ and $\bar{T}_{a}$
correspondingly.) It should be noted that identical intrinsic temperatures of
matter and antimatter $T_{m}=\bar{T}_{a}$ do not ensure equilibrium between
them, since transferring energy from antimatter to matter (from our
perspective, and from matter to antimatter from the perspective of the
antiobserver) increases the overall statistical weight and is strongly
favoured by thermodynamics. If matter and antimatter are to be placed in
conditions of thermodynamic equilibrium, both directions of heat transfer
$\delta Q>0$ and $\delta Q<0$ must be equally likely and have the same
statistical weight $W_{tot}$. This occurs only when $T_{m}=-\bar{T}_{a}$
indicating that the apparent temperature of antimatter is $T_{a}=-\bar{T}_{a}$
so that the conventional equilibrium condition in our frame of reference is
given by $T_{m}=T_{a}$. In the same way, the antiapparent (i.e. perceived by
the antiobserver) temperature of matter is $\bar{T}_{m}=-T_{m}$. It is easy to
see that thermodynamic quantities $\bar{S}_{a},\;\bar{T}_{a},\;$and
$\bar{U}_{a}$ that characterise the intrinsic properties of antimatter are
apparent as%
\begin{equation}
T_{a}=-\bar{T}_{a},\;\;S_{a}=-\bar{S}_{a},\;\;U_{a}=\bar{U}_{a}\label{TSU}%
\end{equation}
from our perspective. The sign of $U_{a}$ is selected to be consistent with
the first law of thermodynamics (\ref{Utot}), while the sign of $S_{a}$ is
chosen to be consistent with the definition of temperature $T_{a}%
^{-1}=\partial S_{a}/\partial U_{a}$\ and with equations (\ref{TT}). The
change of sign does not affect our interpretation of reversible
transformations of antimatter since $S_{a}$ is constant whenever $\bar{S}_{a}$
is constant, which is consistent with our assumption that matter and
antimatter behave in the same way in reversible processes.

It appears that negative temperatures created in our world can, at least in
principle, be placed into thermal equilibrium with thermodynamic antimatter,
which is controlled by reverse causality. If, according to the Boltzmann
hypothesis, the direction of time and causality are determined by the action
of the second law of thermodynamics, then one might expect the existence of a
degree of reverse causality for matter in states with negative temperatures.

\subsection{Apparent heat capacities}

Equation (\ref{TSU}) indicates that thermodynamic considerations of antimatter
involve not only negative temperatures but also negative heat capacities.
Indeed, the equation linking the heat capacity $C$ to $S,U$ and $T$ (for
example, see \cite{K-OTJ2012})
\begin{equation}
C_{a}=\frac{-1}{T_{a}^{2}}\left(  \frac{\partial^{2}S_{a}}{\partial U_{a}^{2}%
}\right)  ^{-1}=\frac{1}{\bar{T}_{a}^{2}}\left(  \frac{\partial^{2}\bar{S}%
_{a}}{\partial\bar{U}_{a}^{2}}\right)  ^{-1}=-\bar{C}_{a}\label{CC}%
\end{equation}
indicates that changing the sign of the entropy $S$ changes the sign of the
heat capacity $C$ irrespective of the sign of the temperature $T$.
Negativeness of antimatter heat capacities is illustrated in Figure
\ref{fig4}. The heat capacity of antimatter is positive according to the
antiobserver and negative according to the observer. As seen by the
antiobserver, two thermodynamic antimatter objects with initial
intrinsic\ temperatures $\bar{T}_{1}$ and $\bar{T}_{2}$ at $\bar{t}=-t_{0}$
are brought into contact. The second object is presumed to be somewhat
intrinsically hotter than the first objects $\bar{T}_{2}>\bar{T}_{1}$ so that
heat $\Delta Q>0$ is transferred, according to an antiobserver, from the
second object to the first object to reach the equilibrium temperatures
$\bar{T}_{0}$ at $\bar{t}=+t_{0}$. Note that $\bar{T}_{2}>\bar{T}_{0}>\bar
{T}_{1}>0$ and $\bar{C}_{1},\bar{C}_{2}>0,$ while the observational symmetry
principle requires that, according to any antiobserver, the thermodynamic
properties of antimatter are the same as thermodynamic properties of matter
observed by us.

The same process looks different from our (the observer's) perspective. The
initial equilibrium state with temperature $T_{0}$ at $t=-t_{0}$ is replaced
by non-equilibrium states $T_{1}>T_{2}$ at $t=+t_{0}$ due to transferring heat
from the first body to the second body. One can easily see that the condition
$T_{1}>T_{2}$ does not allow for positive heat capacities $C_{1}$ and $C_{2}$.
This explains why positive heat capacities $\bar{C}_{1},\bar{C}_{2}>0$ seem
negative $C_{1},C_{2}<0$ to us. The initial equilibrium state detected by the
observer is unstable --- when the heat capacities are negative, a small amount
of heat $\delta Q$ transferred from the first object to the second object
increases the apparent temperature of the former and decreases the apparent
temperature of the latter. This stimulates further apparent transfer of heat
from hotter to colder objects, i.e. in the same direction. Instability of
equilibrium states is a general property of negative heat capacities as
considered in the following section. Note that both the observer and the
antiobserver determine that the process is consistent with the second law as
seen from their respective perspectives --- the heat is transferred from
hotter to colder objects. The possibility of thermodynamic description
backwards in time comes as the cost of introducing unusual ''negative''
thermodynamics. We stress that our prime goal is not in a rather formal task
of interpreting heat transport backward in time but in analysis of a more
complex and interesting problem --- interactions of matter and antimatter.

\subsection{Mass exchange between matter and antimatter}

An antimatter system with a variable number of particles is characterised by
the equation
\begin{equation}
d\bar{U}_{a}=\bar{T}_{a}d\bar{S}_{a}+\bar{\mu}_{a}d\bar{N}_{a}%
\end{equation}
which remains conventional as long as it is presented from the perspective of
the antiobserver. From our perspective, this equation changes according to
(\ref{TSU}). The reversible equivalence requires preservation of mass, which
demands that the apparent and intrinsic numbers of particles composing
antimatter are the same $N_{a}=\bar{N}_{a}$. Hence $\mu_{a}=\bar{\mu}_{a}$.

Consider a reaction converting matter to into antimatter (one-to-one to
preserve the total mass) under conditions when the energy of matter and
antimatter remain the same $dU_{m}=0$ and $dU_{a}=0$. If $dN$ particles have
been converted from antimatter to matter, the change in total apparent entropy
is given by
\begin{equation}
dS_{tot}=dS_{m}+dS_{a}=-\frac{\mu_{m}}{T_{m}}dN_{m}-\frac{\mu_{a}}{T_{a}%
}dN_{a}=-\left(  \frac{\mu_{m}}{T_{m}}+\frac{\bar{\mu}_{a}}{\bar{T}_{a}%
}\right)  dN\label{mu1}%
\end{equation}
where we take into account that $dN_{m}=+dN$ and $d\bar{N}_{a}=dN_{a}=-dN$ and
that $\bar{T}_{a}=-T_{a}$. Assuming that $\bar{T}_{a}=T_{m},$ the
observational symmetry principle requires that $\bar{\mu}_{a}=\mu_{m}$
resulting in
\begin{equation}
dS_{tot}=-2\frac{\mu_{m}}{T_{m}}dN\label{mu2}%
\end{equation}
Hence thermodynamics strongly favours transfer from antimatter to matter if
$\mu_{m}<0$ and vice versa if $\mu_{m}>0$ (according to the observer). In this
case equivalence of the observed chemical potentials $\mu_{a}=\mu_{m}$ does
not ensure equilibrium of the matter/antimatter reaction, since the
corresponding apparent temperatures $T_{a}\neq T_{m}$ are not at equilibrium.
If, however, matter and antimatter are somehow brought into thermal
equilibrium and have the same apparent temperatures $T_{m}=T_{a}$, the
chemical potentials $\mu_{m}$ and $\mu_{a}=\bar{\mu}_{a}$ would not coincide
since matter and antimatter must have very different properties at very
different intrinsic temperatures $T_{m}$ and $\bar{T}_{a}=-T_{m}.$ The
properties of matter and antimatter are determined by their intrinsic
parameters, while equilibrium between matter and antimatter is controlled by
apparent (or antiapparent) values of the parameters. 

\section{''Negative'' thermodynamics}

In this section we briefly review general features of systems with negative
temperatures and negative heat capacities, which are not necessarily related
to the apparent properties of antimatter and have been repeatedly discussed in
the literature for other applications
\cite{Ramsey1956,LL5,Bell1999,sodium147,K-OTJ2012}. This section is presented
from the observers perspective.

\subsection{Negative temperatures}

In this section we discuss general properties of negative temperatures, which
may or may not be related to reversal of the direction of time and
matter/antimatter interactions. In our world, negative temperatures can appear
in systems with a limited number of microstates at high energies
\cite{Ramsey1956}. A very simple example of a system with negative
temperatures can be found in Ref. \cite{K-OTJ2012} The effects encountered in
lasers and, possibly, in biological organisms bear resemblance to negative
temperatures. Practically, negative temperatures are difficult to create and
even more difficult to maintain since systems with negative temperatures tend
to be unstable \cite{LL5}.

Unlike the mathematical convention of ordering positive and negative numbers,
negative temperatures are not lower but higher than the positive temperatures
\cite{Ramsey1956,K-OTJ2012}. The lowest possible temperature is $T=+0.$
Positive temperatures can increase up to $T=+\infty,$ which is the same as
$T=-\infty$ and can be interpreted as the effective temperature of work.
Negative temperatures can increase further up to $T=-0$, which is the highest
possible temperature. Is is difficult to remove heat from low temperatures and
add heat to high temperatures. One can see that temperature ordering is
consistent with conventional mathematical ordering of the following quantity
\begin{equation}
\beta=-\frac{1}{T} \label{bet}%
\end{equation}
which we can call quality of energy or\ quality of heat. The quality of work
corresponds to $\beta=0$. Conventional heat with positive temperatures is of
lower quality $\beta<0,$ while heat with negative temperatures is of higher
quality $\beta>0$ than work. Note that infinite temperatures do not
necessarily need to posses infinite energies and can be linked to a loss of
the dependence of the number of stochastic degrees of freedom present in the
system on energy. Energy without a random component, i.g. work or coherent
light, are assigned the values of $\beta=0$ and $T=\infty.$

The second law of thermodynamics indicates that the overall quality of heat
cannot increase in an isolated system. While the quality of heat can decrease
in an isolated system, this process is irreversible as subsequent increases of
the heat quality are not permitted. For example, as shown in Figure
\ref{fig2}(a) transition of heat $\delta Q$ from $\beta_{2}$ to $\beta_{1}$ is
possible but it is irreversible since transition $\delta Q$ from $\beta_{1}$
to $\beta_{2}$ is prohibited by the second law. Any reversible change of heat
quality needs to involve at least three different levels to preserve the
overall quality of heat. The reversibility condition for the process shown in
Figure \ref{fig2}(b) is thus%
\begin{equation}
\beta_{0}\left(  \delta Q_{H}+\delta Q_{L}\right)  =\beta_{H}\delta
Q_{H}+\beta_{L}\delta Q_{L} \label{betLH}%
\end{equation}
that is the increase of the heat quality from $\beta_{0}$ to $\beta_{H}$ must
be compensated by the decrease of the heat quality from $\beta_{0}$ to
$\beta_{L}$. Equation (\ref{betLH}) preserves the overall entropy and is valid
for the whole range of $-\infty<\beta<+\infty$.

Any engine is designed to produce work, hence one of its energy qualities that
corresponds to the produced work is $\beta=0$. A hypothetical reversible
engine producing work in most efficient way allowed by the second law is
conventionally referred to as Carnot engine. As a reversible engine, Carnot
engine must\ comply with condition (\ref{betLH}). There are three modes that a
Carnot engine can operate in, as shown in Figure \ref{fig3}. The first is the
conventional Carnot engine working with positive temperatures, where the
quality of heat $\delta Q_{H1}$ is increased from $\beta_{H1}<0$ to work
$\beta=0$ with compensating decrease of the quality of heat $\delta Q_{L1}$
from $\beta_{H1}$ to $\beta_{L1}$. The second engine works with negative
temperatures: the quality of heat $\delta Q_{L2}$ is decreased from
$\beta_{L2}>0$ to work $\beta=0$ which gives the opportunity to increase
quality of some of the heat $\delta Q_{L2}$ from $\beta_{L2}$ to $\beta_{H2}$.
The third engine works with one positive and one negative temperature heat
reservoirs and manages to produce work from both of these sources. We stress
that this does not contradict the second law since the heat quality increases
from $\beta_{L3}$ to $\beta=0$ and decreases from $\beta_{H3}$ to $\beta=0$
while the overall quality of energy remains the same.

\subsection{Negative heat capacities}

Although not common in conventional thermodynamics, negative heat capacities
have been repeatedly discussed in the literature
\cite{Bell1999,sodium147,K-OTJ2012}. From the point of view of conventional
thermodynamics, negative heat capacities $C<0$ are even more unusual than
negative temperatures $T<0$. \ A thermodynamic system with negative $C$ can
not be divided into equilibrated subsystems and its thermodynamics can not be
evaluated with the use of partition functions \cite{K-OTJ2012}. Indeed, if a
system consists of two equilibrated systems 1 and 2 with $C_{1}<0,$ $C_{2}<0$
and $T_{1}=T_{2},$ this equilibrium is unstable since transferring $\Delta
Q>0$ from system 2 to system 1 increases $T_{2}$ and decreases $T_{1},$ which
further stimulates heat transfer in the same direction. Stars and black holes
may serve as realistic examples of systems with negative heat capacities
\cite{Bell1999,K-OTJ2012}.

Interactions of systems with negative and positive heat capacities are capable
of unusual behaviour that seemingly violates the second law of thermodynamics,
but in fact the average quality of energy in these interactions stays the same
or decreases and the entropy does not decrease. Consider the example shown in
Figure \ref{fig5}. Interactions of two systems with $-C_{2}=C_{1}>0$ may
result in an apparent increase of their temperatures through reversible hear
transfer $\Delta Q$ occurring between the same temperatures. Note that,
according to the definition of heat capacity
\begin{equation}
\frac{\partial S}{\partial T}=\frac{C}{T}%
\end{equation}
the entropy $S_{2}$ decreases in this process by the same amount as the
entropy $S_{1}$ increases (assuming $T>0$) and the overall quality of thermal
energy remains the same. Since the process is reversible when the absolute
values of the heat capacities are the same, apparent decrease in the
temperatures due to negative $\Delta Q$ is also possible. If, however,
$\left|  C_{2}\right|  <\left|  C_{1}\right|  $, the process of temperature
increase becomes irreversible since $T_{2}$ grows faster than $T_{1}$ as shown
in Figure \ref{fig5}. In this case the overall entropy $S_{1}+S_{2}$ increases
and some of the quality of heat is lost.

\section{Examples}

This section offers a number of gedanken experiments that are designed to
illustrate the properties of thermodynamic antimatter. While the overall setup
of these experiments is, of course, unrealistic, the points illustrated by
them may well be quite relevant to the real world.

\subsection{On shaking hands with anti-people}

First we recall the well-known warning of Richard Feynman (based on CP
invariance) not to shake alien left hand if this hand is offered for the
handshake --- the alien might be made from antimatter. We act cautiously and
decide to send a robot to perform this handshaking mission. Everything begins
from a radio message arriving from the outer space thanking us for sending our
robotic representative to shake hands with their robotic representatives at a
given very remote location --- this encounter was most educational. We look at
that location through a telescope and see what seem to be remaining of a big
explosion but still decide to send a robot there. The robot is instructed to
offer a handshake at certain location in space and time. As our robot
stretches his robotic arm, another robot (an antirobot, as the reader has
already guessed) materialises from the surrounding dust cloud and, for an
instant, shakes the stretched hand (Figure \ref{fig7}). In accordance with the
thermodynamics of antimatter, our robot receives a very large amount of energy
from the antirobot, which blows him into small pieces. This effect is purely
thermodynamic (the same as shown in Figure \ref{fig1}) --- it does not need
annihilation of physical matter and antimatter occurring on their contact. The
alien antirobot then retreats backwards into his alien antiworld, while we do
not have any realistic chance of reassembling our robot back. As polite
people, we do not forget to send a radio message thanking our counterparts for
the educational encounter.

Could this meeting be more productive without shaking hands? It certainly
could, but we must keep in mind that many other accidents are
thermodynamically possible during the meeting. For example, as the robots
approach each other, they might be hit by a coherent light beam. While our
robot reports being hit by the beam radiated by the alien robot, his
counterpart makes exactly the same compliant. These complaints miss the main
point that the beam converting energy from negative temperatures ($\beta>0$)
into coherent radiation ($\beta=0$) and from coherent radiation into positive
temperatures ($\beta<0$) is strongly favoured by thermodynamics. Whether the
beam actually occurs or not is determined by the kinetics of the process and,
as all kinetic issues, is outside the scope of our analysis. 

\subsection{Travelling to the antiworld}

A prudent traveller to distant places must take enough fuel to provide a
continuous supply of energy but this is going to be the least difficult part
for our planned journey. Any object found in the antiworld would represent a
source of energy of very high quality. Here we do not need to interpret
antimatter as composed of antiprotons and positrons, which can readily react
with protons and electrons to release plenty of energy (although this indeed
might be the case). The energy is provided thermodynamically by antimatter
releasing large amounts of energy as soon as we come into a thermodynamic
contact with it. This indicates that the antiworld is a rather dangerous place
for us. Specifically we should keep away from any objects we might see there,
anti-stars and anti-planets (one may note that anti-planets are hotter for us
than anti-stars). Antiobservers populating the antiworld would see us as
excessively hot and dangerous.

The question if we would be able see anything in the antiworld is reasonable
question to ask. The temperatures of anti-stars $T\approx-6000K$ is well below
that of background radiation (combined with some dispersed antimatter), which
is extremely hot at $T\approx-3K$ (if we use the familiar conditions from our
world as a model for the antiworld). Thus transfer of radiative energy is
directed from background to the stars. The hottest object we may find in the
antiworld are black holes, which we see as white holes due to reversal of
time. The temperature $T$ of large white holes approaches $-0K$ and almost
nothing can enter its horizon. An antimatter object at $T\approx-3K$, although
very hot, may be less dangerous for us than another ''cooler'' object at, say
$T\approx-6000K,$ since the first object may have very little energy left to
pass onto us during a contact (while the object keeps increasing its
temperature from $T\approx-3K$ to $T\approx-0K)$. 

The most interesting part in our journey is, of course, not interpreting
conventional thermodynamics backward in time but the interactions of
antimatter (antiworld) and matter (us). The prospects of this interaction are
not particularly soothing. The largest problem we will have to face is not
lack of energy but an excess of it. The antiworld around us is so hot that we
will not be able to dump excess heat anywhere. The only opportunity we have is
delaying our imminent destruction due to overheating by having a perfectly
reflecting coating on our space ship and taking plenty of ice for cold drinks
with us. If we stay away from antimatter and its radiation, we might be lucky
to stretch our journey little bit longer in our time.

In any case, we conclude that the antiworld is a very hostile environment for
matter and traveling there is not recommended. The same travel warning applies
to any attempts to build a time machine that allows us to move back in time
but is not capable of complete thermodynamic separation of the time machine
from the rest of the world. Note that matter moving back in time acquires
thermodynamic properties of antimatter. We should not blame the antiworld for
its hostile attitude towards us as our world is no less hostile towards
visitors from antiworld and an anti-traveller would encounter the same
problems in our world as we encounter in his. In the conditions currently
prevailing in the universe, the opportunities for existence of macroscopic
quantities of thermodynamic antimatter are very limited.

\subsection{''Thermodynamic Bang''}

The process of having thermodynamical matter, antimatter and radiation in
equilibrium at infinite temperature with subsequent cooling is called here the
thermodynamic bang. This, of course is not a model for the real Big Bang but
an illustration of direction of the processes that may thermodynamically occur
under specified conditions (as indicated by reactions I, II and III in Figure
\ref{fig6}).The division of the cooling process into stages is purely
schematic and given as illustration for the direction of the processes for
each of the reaction, while overall evolution of the system is determined by
reaction kinetics with all three reactions occurring at the same time.

\textbf{The bang}. The point with $T_{m}=\bar{T}_{a}=\infty=T_{a}$ (i.e.
$\beta_{m}=\bar{\beta}_{a}=0=\beta_{a}$ where $\beta=-1/T)$ is assumed to be
located at $t=\bar{t}=0$ as shown in Figure \ref{fig6}. When the temperatures
are close to infinity, the system is dominated by radiation but may include
matter and antimatter in the state of thermodynamic equilibrium. The chemical
potentials of radiation, matter and antimatter are conventionally set to zeros.

\textbf{Stage I.} As the system cools adiabatically due to expansion in the
direction of positive $t$, matter $\frak{m}$, antimatter $\frak{a}$ and
radiation $\frak{r}$ can react according to
\begin{equation}
\text{(I) }\frak{r}\rightleftarrows\frak{m}+\frak{a}%
\end{equation}
producing more matter and antimatter from radiation. Assuming that adiabatic
expansion is reversible and that there are no thermodynamic interactions
between matter and antimatter, the intrinsic temperatures of matter and
antimatter must be the same $T_{m}=\bar{T}_{a}$ (that is $\beta_{m}=\bar
{\beta}_{a}$) and decreasing. This corresponds to negative apparent
temperatures of antimatter $T_{a}=-\bar{T}_{a}<0$ (and $1/\beta_{a}%
=-1/\bar{\beta}_{a}>0$). Under these conditions reaction (I) can occur
reversibly as illustrated in Figure \ref{fig2}(b), although produced matter
and antimatter are not in thermal equilibrium with each other since they have
different apparent temperatures $T_{m}$ and $T_{a}=-T_{m}$. The temperature of
the radiation, which can not be equilibrated by matter or antimatter, remains
effectively infinite (if the notion of temperature can be applied to radiation
at this stage). The chemical potentials of matter and antimatter are the same
$\mu_{m}=\mu_{a}=\bar{\mu}_{a}$ due to the observational symmetry but do not
have to take zero values since reaction (I) is not at its thermodynamic
equilibrium and the condition $\mu_{r}=$ $\mu_{m}+\mu_{a}$ does not apply.
Expansion and loss of pressure is likely to make $\mu_{m}=\mu_{a}=\bar{\mu
}_{a}$ negative.

\textbf{Stage II.} The thermodynamic interactions of matter and antimatter
result in heat being transferred from antimatter to matter according to the
equation
\begin{equation}
\text{(II) }\left(  U_{a}+\Delta Q\right)  +U_{m}\rightleftarrows
U_{a}+\left(  U_{m}+\Delta Q\right)
\end{equation}
reheating matter and, possibly, stimulating inflational expansion. The
apparent temperatures increase similar to the mechanism shown in Figure
\ref{fig5}. The heat transfer specified by (II) is irreversible since
$T_{a}>T_{m}$.

\textbf{Stage III.} If the following reaction is allowed
\begin{equation}
\text{(III) }\frak{a}\rightleftarrows\frak{m}%
\end{equation}
antimatter is then converted into matter (and, possibly, also annihilate with
matter into radiation according to reaction (I)). Note that reaction should
preserve the intrinsic temperature $T_{m}^{\circ}=\bar{T}_{a}$ due to
conservation of energy and similarity of matter and antimatter. Hence reaction
(II) acts to\ cool matter and the rest of the system since the temperature
$T_{m}^{\circ}$ of the converted matter is below the temperature $T_{m}$ of
the existing matter. Assuming negative chemical potential of  matter, reaction
(III) increases the entropy due to equation (\ref{mu2}). Both reactions (II)
and (III) are thus irreversible and predominantly proceed in the forward
direction. Finally, as antimatter disappears, interactions of matter and
radiation establish equilibrium between them. From\ this moment, conventional
forward-time causality is established and antimatter may exist only as a
thermodynamic fluctuation.

The same process may occur backward in time establishing dominance of
antimatter and inverse causality. Thus the antiworld is located ''before'' the
bang in our time and there is no way for us to access it. According to the
observational symmetry principle, anti-people populating the antiworld would
say exactly the same things about us. Disappearance of antimatter at $t>0$
converts the initial time-elliptic equilibrium state into highly
non-equilibrium state with directional time. The stated negativeness of the
chemical potentials $\mu_{m}=\mu_{a}=\bar{\mu}_{a}$ is essential for the
suggested scheme. If we assume $\mu_{m}=\mu_{a}=\bar{\mu}_{a}>0$ antimatter is
going to be dominant at $t>0$ and matter is going to be dominant at $t<0,$
then our ''thermodynamic bang'' becomes a ''thermodynamic crunch''.

In our analysis, it is important that the time coordinate is not looped,
although closed time-like loops are formally possible in general relativity.
Time loops are likely to stimulate antimatter production. Indeed, if time
forms a loop, then the production of matter in one direction is not separated
from the production of antimatter in the opposite direction. A prudent space
traveller should thus beware of any temporal Merry-go-round if he does not
want to experience an ''antimatter shower''.

The assumptions made in this paper necessitate that, in absence of matter and
antimatter, radiation can not change its entropy. In our world, interactions
of radiation and matter result in increase of the overall entropy of matter
and radiation forward in time. In the same way, interactions of radiation and
antimatter in absence of matter result in increase of the overall entropy
backward in time. If both matter and antimatter are present in large
quantities, radiation should have infinite temperature (i.e. stay coherent).
We also should note that no black/white holes can be formed in this mixture.
Indeed, matter forms black holes while antimatter forms white holes due to
corresponding forward and backward irreversibilities of these processes. Can a
hole be black and white at the same time? The exterior of a black hole is no
different from that of a white hole. Inside the horizon, both holes have the
same metric but with a different direction of the time-like coordinate: the
singularity is in the future of black holes and in the past of white holes.
Thus, a hole may switch from white to black but can not be white and black at
the same time (i.e. under the same horizon). Hence, the mixture of matter and
antimatter can not form black/white holes even if matter and antimatter
gravitationally attract each other. This might be the possible driving factor
that ensures perfect uniformity of the bang --- it is just cannot exist in any
other way. The time is non-directional near the bang point $t\sim0,$ which
means by definition that there are no white or black holes since these holes
are time-directional. 

\section{Discussion}

While this work was started as an entertaining thermodynamic exercise, it
seems now that it has implications stretching beyond our original intentions.
The presented consideration connects the perceived direction of time to
abundance of matter in the universe, to thermodynamic antagonism of matter and
antimatter and to the fact that, as far as we know, time does not form a
closed loop. Although it is always possible to justify the arrow of time by
introducing causality, this corresponds to postulating direction of time
without explaining its physical mechanism \cite{PriceBook,K-PS2012T} (and
there must be one).

The Boltzmann time hypothesis is enlightening in many respect and the temporal
boundary conditions imposed on the universe are important but, when taken
alone, insufficient to explain why causality is routinely observed at almost
every scale. Matter should have at least some time-directional bias, which
must be very small as we do not detect it in both classical mechanics and
conventional quantum mechanics. The bias, nevertheless, has a profound effect
on the universe. Penrose \cite{PenroseBook} provides with a number of
convincing arguments supporting this view but, ultimately, the exact physical
mechanism of running time is yet to be found. Considering that, to the best of
our knowledge, the world is CPT-symmetric and very close to CP-symmetric,
matter should be in a minor violation of the time symmetry. The apparent
problem comes from antimatter and from the Feynman theory \cite{Feynman1949},
which treats antiparticles as moving back in time. If matter has a slight
forward time bias, the CPT symmetry requires  antimatter to have the same bias
backward in time. How this can be consistent with the conventional second
law\cite{CPT2012} ? Our CPT-invariant thermodynamics seems to resolve this
issue. We stress, however, that this paper introduces a consistent
thermodynamic theory constructed on postulated properties of thermodynamic
antimatter, while the degree of its applicability to real antimatter is yet to
be determined. This investigation is by far not trivial since the exact
mechanism enacting the direction of time is not known and must be very fine
while intrinsic thermodynamic properties of small quantities of antimatter can
be overpowered by interactions with matter--populated outside world, even if
these interactions remain very weak. At present, the apparent absence of large
quantities of antimatter in the universe gives the best support to the theory.

In absence of antimatter, the CPT-invariant thermodynamics should coincide
with conventional thermodynamics and, generally, this is the case. Has any
conventional thermodynamic property been lost in the introduced
thermodynamics? Yes, there is one --- the photon gas (which is the same as
anti-photon gas) has lost its ability to independently comply with the second
law. Consider gedanken experiment suggested by Hawking \cite{Hawking1976}:
either radiation or matter and radiation are placed in a box (which may reach
a galactic scale and contain black holes) with perfectly reflective insulating
boundaries (mirrors). There is an expectation that a mixture of matter and
radiation will eventually come into equilibrium. Our thermodynamics agrees
with this. However, if radiation is present in the box without matter, its
evolution according to our theory becomes time reversible, its entropy does
not increase and the gas can not be equilibrated. Is this a problem? We
would\ argue that it is not, since either the box mirrors are made of matter
and are not perfectly reflective or this experiment can hardly be conducted in
any practical form or shape. Thermodynamic equilibration of the pure photon
gas in a universe, which is free of any matter and antimatter, is not an
experimental fact but a hypothesis primarily based on validity of the
conventional second law of thermodynamics.

\section{Conclusions}

The present work follows modern understanding of the Boltzmann time hypothesis
and Feynman concept of antimatter as being similar to matter but moving
backward in time. The developed thermodynamics of matter/antimatter
interactions is both CPT-invariant and irreversible at the same time. The
forward-time irreversibility is seen as the property of matter populating the
world, which is accompanied (at least in principle) by backward-time
irreversibility of thermodynamic antimatter. Matter and antimatter, however,
display the same properties in reversible processes. Our direction-neutral
consideration of thermodynamics is necessarily accompanied by suspending the
causality principle. The analysis of thermodynamic interactions of matter and
antimatter leads to negative temperatures: the temperature of antimatter,
which is determined by the antiobserver as being positive, seems negative to
us. Negative temperatures make prolonged interactions of matter and antimatter
impossible and the paucity of antimatter in our world is thus not surprising.
The thermodynamic interactions of matter and antimatter have a profound
influence on the perceived direction of time, on dominance of matter over
antimatter and on unique properties of the universe at its origin.

\begin{center}

\begin{figure}[tbp]
\includegraphics[width=14cm]{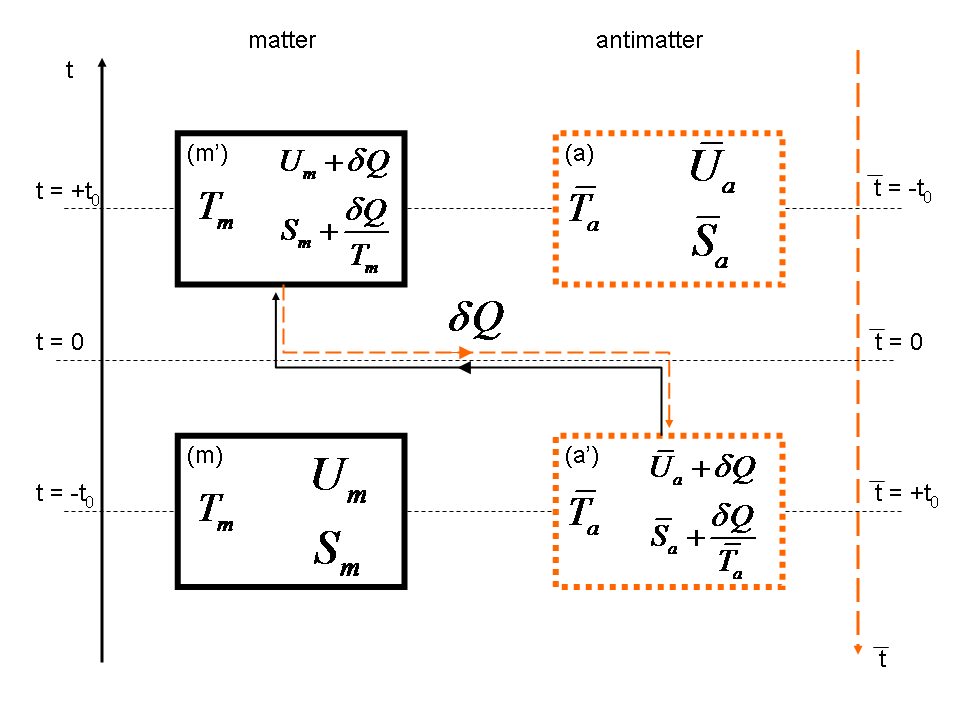}
\caption{Thermodynamic interaction of matter and animatter.}
\label{fig1}
\end{figure}

\begin{figure}[tbp]
\includegraphics[width=14cm]{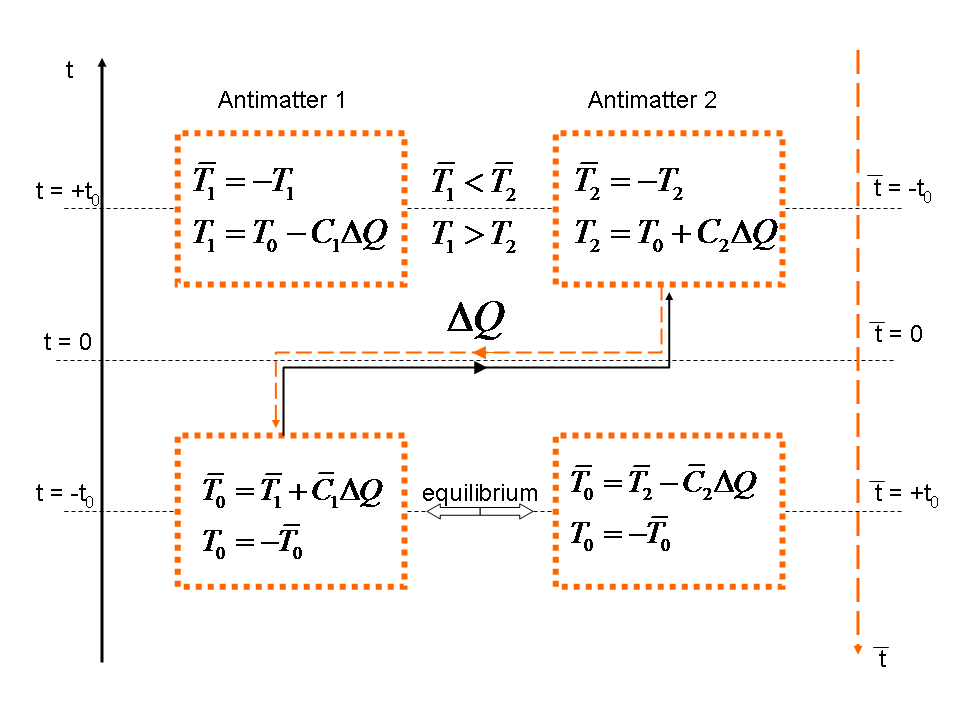}
\caption{Antimatter heat transfer as seen by antiobserver and by us.}
\label{fig4}
\end{figure}

\begin{figure}[tbp]
\includegraphics[width=14cm]{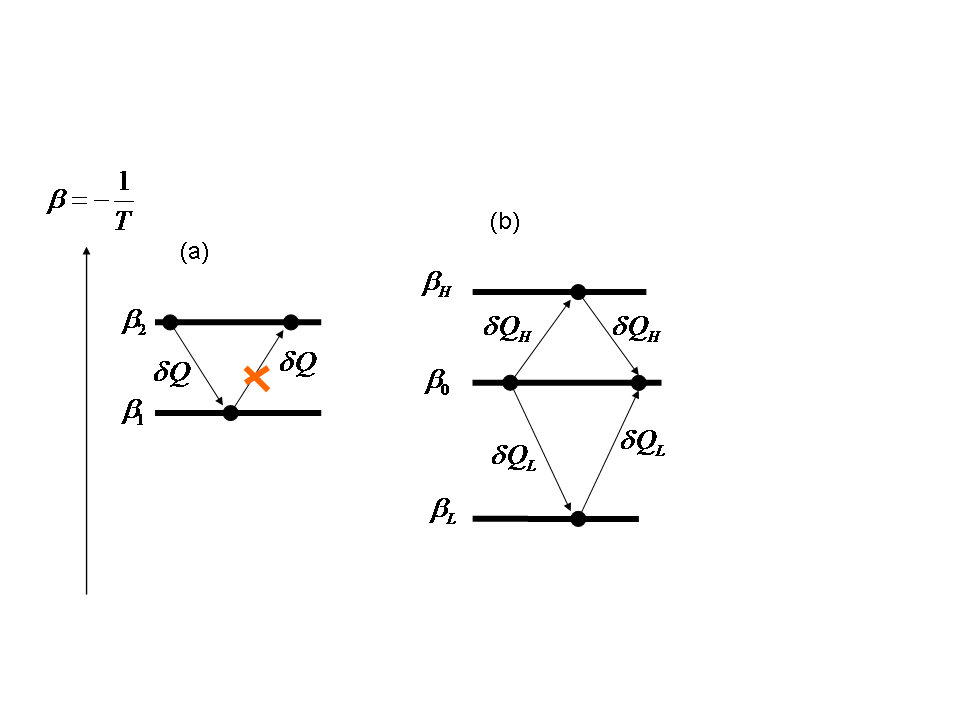}
\caption{Change in qualities of energy a) irreversible and impossible, b) reversible if the overall quality of energy is preserved.}
\label{fig2}
\end{figure}

\begin{figure}[tbp]
\includegraphics[width=14cm]{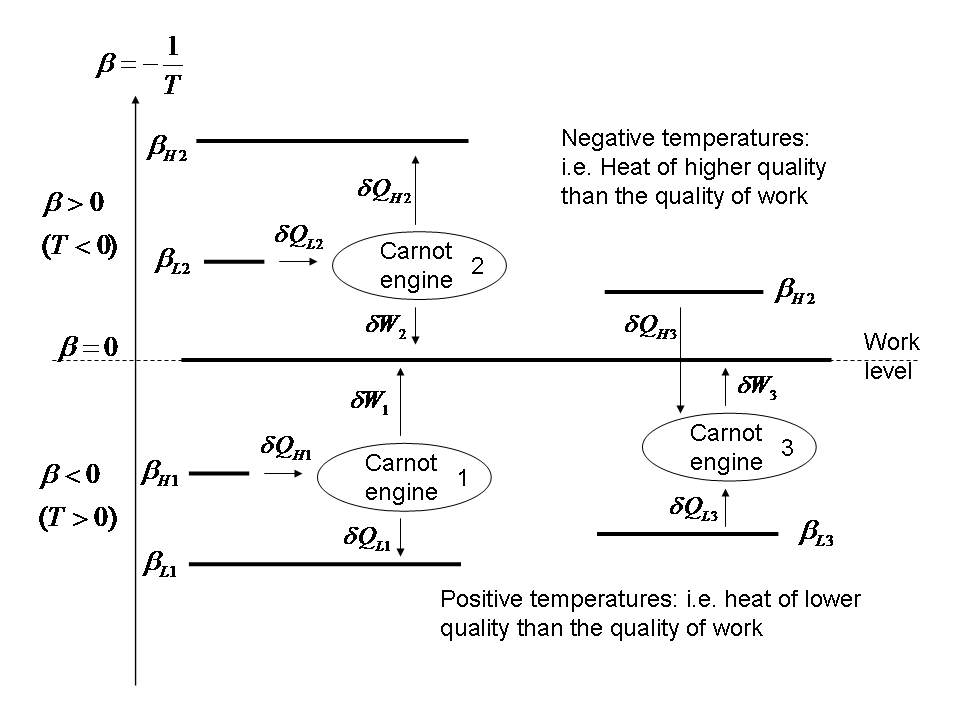}
\caption{Three regimes for ideal reversible (Carnot) engine.}
\label{fig3}
\end{figure}

\begin{figure}[tbp]
\includegraphics[width=14cm]{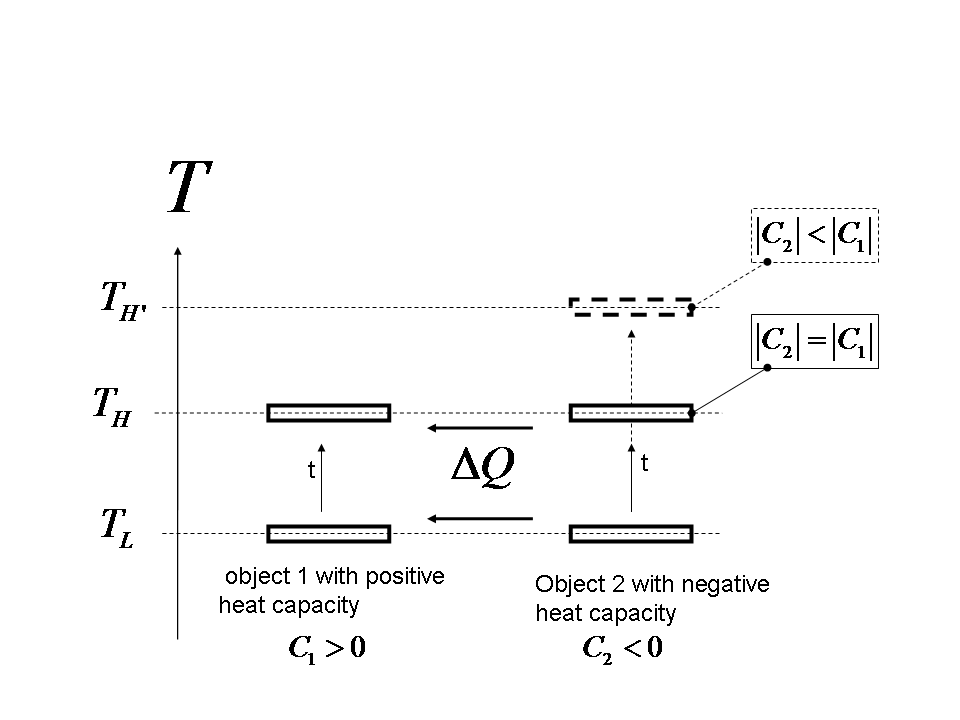}
\caption{Heat transfer when heat capacity might be negative. }
\label{fig5}
\end{figure}

\begin{figure}[tbp]
\includegraphics[width=14cm]{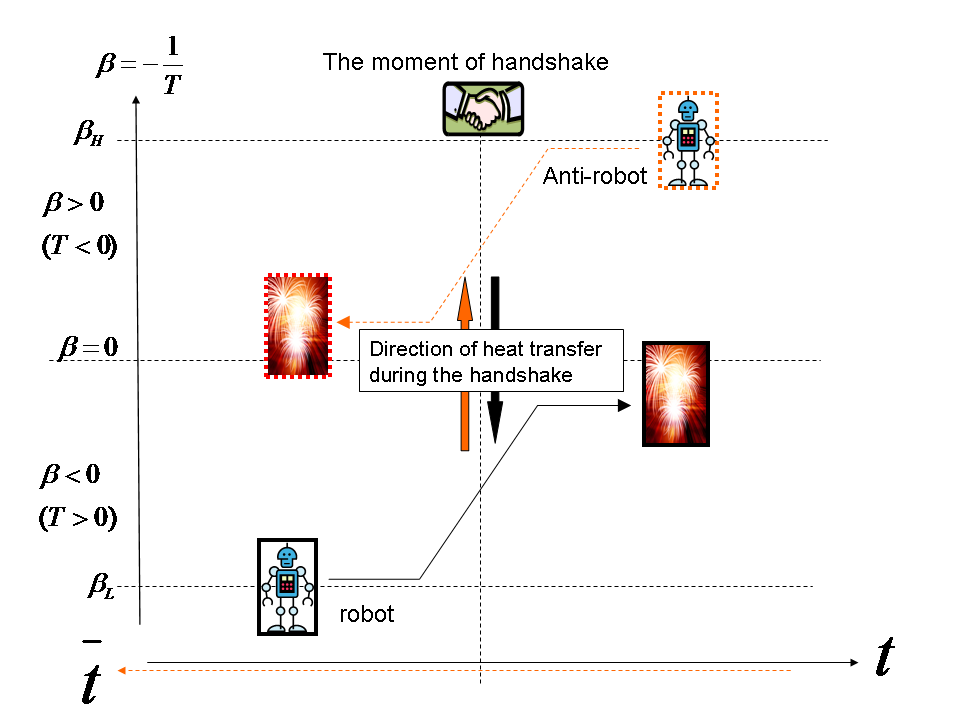}
\caption{Robot and antirobot shake hands.}
\label{fig7}
\end{figure}

\begin{figure}[tbp]
\includegraphics[width=14cm]{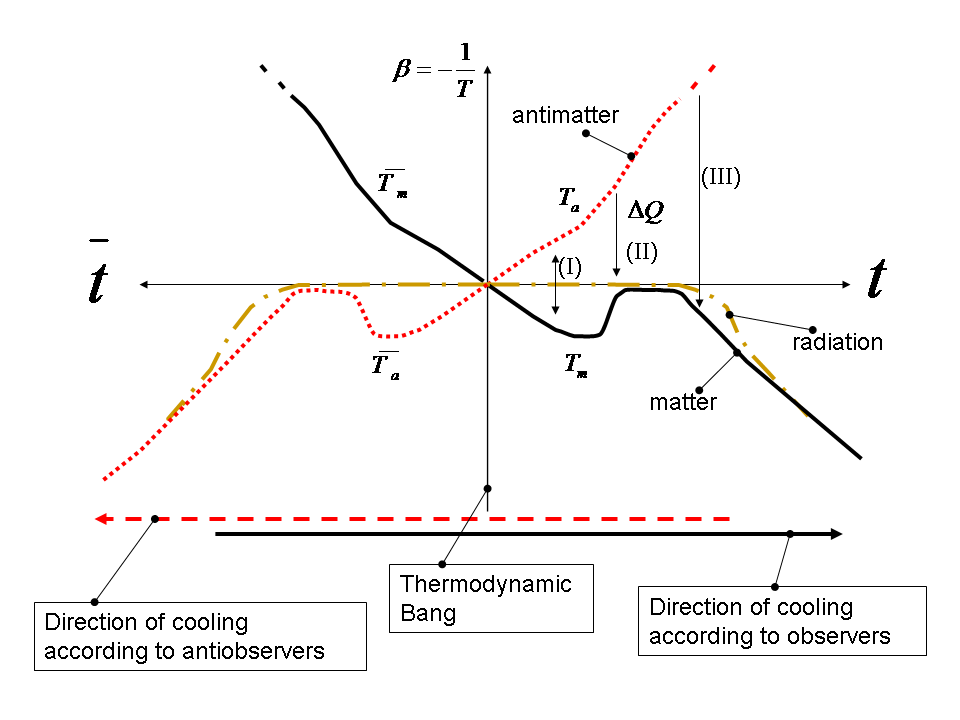}
\caption{Thermodynamic Bang}
\label{fig6}
\end{figure}

\end{center}
\end{document}